\documentclass[amsmath, amssymb, aps, physrev, 10pt]{revtex4-2}

\usepackage{graphicx}
\usepackage{dcolumn}
\usepackage{bm}
\usepackage{dsfont}
\usepackage{amsmath}
\usepackage{color}
\usepackage{MnSymbol}
\usepackage{subfigure}
\usepackage{ragged2e}
\usepackage{float}
\usepackage{silence}
\def\d{\text{d}}
\def\M11{$\mathbb M^{1+1}$}
\def\orderof#1{\mathcal{O}(#1)}
\def\quabla{\Box}
\def\precstar{\prec\hspace{-1.4mm}\star}
\def\CCC{\mathbf{C}}
\def\LLL{\mathbf{L}}
\def\GGG{\mathbf{G}}
\def\DDD{\mathbf{D}}
\def\KKK{\mathbf{K}}
\def\SSS{\mathbf{S}}
\def\id{\mathds{1}} 

\begin{document}

\title{\textbf{Link-based causal set propagators in $1+1$ dimensions}}
\author{Haye Hinrichsen}
\author{Arsim Kastrati}
\email{Contact author: arsim.kastrati@uni-wuerzburg.de}
\affiliation{Faculty for Physics and Astronomy, Julius Maximilians University Würzburg, Am Hubland, 97074 Würzburg, Germany}
\date{\today}


\begin{abstract}
We investigate whether retarded scalar propagators on causal sets can be expressed in terms of the link matrix $\mathbf{L}$. For Poisson sprinklings into 1+1 dimensional Minkowski spacetime, we show by asymptotic analysis and supporting numerical simulations that the averaged massless retarded propagator is naturally associated with a normalized exponential exp$(\mathbf{L})$. We then extend the construction to the massive case via the usual mass-scattering series and obtain good agreement with the continuum propagator after averaging. Finally, we discuss the inverse kernel exp$(-\mathbf{L})$ as a possible candidate for a discrete d’Alembertian. 
\end{abstract}


\maketitle
\parskip 2mm

\section{Introduction}

Since the advent of general relativity, space and time are no longer regarded as passive containers but as a dynamical physical entity modeled by a differentiable Lorentzian manifold. However, there are strong indications that the concept of a smooth spacetime continuum cannot be the final microscopic description of Nature. Perturbative quantum gravity, for example, encounters non-renormalizable ultraviolet divergences, and there are, in fact, many indications suggesting the existence of a minimal resolvable length scale~\cite{garay1995quantum,hossenfelder2013minimal}. A natural candidate is the Planck length 
\begin{equation}
\ell_P=\sqrt{\hbar G/c^3}\approx 1.6\times 10^{-35}\,\text{m},
\notag
\end{equation}
which plays a fundamental role in black-hole thermodynamics and the holographic principle: the Bekenstein-Hawking entropy scales with the horizon area in Planck units, hinting that the number of underlying degrees of freedom may scale with area rather than volume~\cite{bekenstein1973black,bousso2002holographic}. These observations motivate the idea that spacetime may cease to be smooth at the shortest distances and instead possess a potentially discrete or otherwise non-classical microstructure. 

However, how this structure is actually realized is still unknown, and different approaches -- notably string theory, loop quantum gravity, and holographic concepts -- offer different scenarios with different implications~\cite{blau2009string,rovelli2008loop,ammon2015gauge}. Another interesting approach, on which we will focus in the present article, is that of causal sets~\cite{bombelli1987space,surya2019causal}. In causal set theory (CST), spacetime is modeled as a discrete set of events. These events do not possess coordinates or metric distances, but feature only a binary relation, specifying whether pairs of events can interact causally or not. According to the so-called \textit{“Hauptvermutung”} (main conjecture) \cite{Hawking:1976fe,dowker2005causal,muller2025hauptvermutung}, this binary relation alone could be sufficient to encode essentially all emergent metric properties on large scales. In this sense, the macroscopically observed smooth structure of space and time is merely an apparent emergent phenomenon based on a highly disordered discrete microstructure on the Planck scale.

More specifically, a causal set is a countable set $\mathcal C$ of indexable events $c_1, c_2, \ldots$ equipped with a binary relation $\prec$ that specifies whether an event can causally interact with another event or not, thus defining a partial order on $\mathcal C$. This relation defines the causal matrix $\CCC$ with the matrix elements\footnote{Throughout this article, matrices defined on discrete causal sets like $\LLL$ are marked by bold letters.}
\begin{equation}
\label{Cdef}
\CCC_{ij} \;=\; 
\delta(c_i \prec c_j) \;=\;
\begin{cases}
1 & \text{ if } c_i \prec c_j \\
0 & \text{ otherwise. }
\end{cases}
\end{equation}
It is useful to enumerate the events in a natural way such that $\CCC$ takes the form of a strict upper triangular matrix~\cite{rideout1999classical}, reflecting the property of partial ordering.

In addition to causal connections, the so-called \textit{links} play a crucial role in the theory of causal sets. A link is defined as a causal connection that cannot be decomposed into a concatenation of other causal connections. The corresponding link relation, usually denoted by the symbol $\precstar$, therefore introduces the notion of adjacency in a causal set. The associated link matrix $\LLL$ is defined by the matrix elements
\begin{equation}
\label{Ldef}
\LLL_{ij} \;=\;
\delta(c_i \precstar c_j) \;=\;
\begin{cases}
1 & \text{ if } \CCC_{ij}=1 \wedge (\CCC^2)_{ij}=0\\
0 & \text{ otherwise.}
\end{cases}
\end{equation}
Since the link relation $\precstar$ determines the causal relation $\prec$ by transitive closure, it can be regarded as encoding the same underlying order-theoretic information. In the spirit of the aforementioned \textit{Hauptvermutung}, this suggests that the continuum spacetime geometry should be recoverable from $\LLL$ together with the density of elements.

The ultimate goal of a discrete theory of spacetime would be to endow the causal set with an intrinsic dynamics through which the causal network constructs itself, thereby explaining how space and time arise from within themselves. Although there are already various interesting approaches in this direction~\cite{rideout1999classical,ahmed2010indications}, we are still far from such a comprehensive theory. Meanwhile, as a workaround, one uses so-called \textit{sprinklings}. In a sprinkling, point-like events are randomly distributed on a part of a given Lorentzian manifold according to a Poisson distribution with a mean density proportional to the local volume element~\cite{bombelli1987space,dowker2004quantum}. Compared to isometric lattices, this construction has the decisive advantage that it is automatically statistically invariant under the isometries of the manifold \cite{surya2019causal}. After generating and indexing the events, their causal relationship is read off and the matrices $\CCC$ and $\LLL$ are constructed accordingly. Subsequently, one attempts to derive physically relevant quantities on the causal set exclusively on the basis of these matrices, that is, without using metrics or coordinates. These quantities are then averaged over many independent sprinklings and compared with the known continuum results in the limit of large sprinkling density.

To study physical phenomena in causal networks, it is customary to introduce additional fields by assigning one or more field variables to each event. The simplest case is that of a classical scalar field, in which each event $c_i$ is associated with a scalar variable $\phi_i\in\mathbb R,\mathbb C$. These local fields interact along adjacent causal links, leading to a discrete analog of the well-known path integral, which can then be used to calculate discrete field propagators~\cite{johnston2008particle,shuman2024path}. This approach, known as the \textit{path sum method}, works well for massive scalar fields in sprinklings embedded in flat Minkowski spacetimes, as well as in spacetimes with constant curvature~\cite{Kastrati:2025etw}. In particular, in the limit of large distances or, equivalently, high sprinkling densities, and when averaging over many independent sprinklings, one successfully recovers the expected results in the continuum case.

The path-sum construction of causal-set propagators has proved to be very effective, but it also leaves an important conceptual issue open. In practice, the massive propagator is obtained from a \textit{chosen} massless propagator by a discrete analog of the usual mass-scattering expansion. The nontrivial step is therefore the actual choice of the massless propagator itself. In the existing constructions, this choice is guided by the known continuum form of the retarded Green function in the respective embedding dimension: one selects a suitable causal set quantity whose expectation value over many independent sprinklings just reproduces the desired continuum behavior and then builds the massive propagator from it. Although this prescription is natural and successful, it is not derived from a dimension-independent intrinsic principle on the causal set. It is precisely this point that motivates the present work.

Let us illustrate this procedure in the simplest example, namely, the propagator of a classical real-valued scalar field with mass $m$ in Minkowski space \M11. In the continuum case, the retarded propagator $G_m^R$ of such a field obeys the inhomogeneous Klein-Gordon equation
\begin{equation}
\label{MassiveEquation}
\bigl(\quabla - m^2\bigr) G_m^R(t,x)\;=\;-\delta(t) \delta(x)\,,
\end{equation}
where $\quabla=\partial_\mu\partial^\mu=\partial_x^2-\partial_t^2$ is the d'Alembert operator using the mostly-plus convention. In the massless case $m=0$, this reduces to the inhomogeneous wave equation $\quabla G_0^R(t,x)=-\delta(t)\delta(x)$ with the explicit solution~\cite{johnston2008particle}
\begin{equation}
\label{continuumprop}
G^R_0(t,x)\;=\;
\frac12 \theta(t) \theta(t^2-x^2)\,.
\end{equation}
In other words, in the continuum case, the retarded massless propagator is constant and equal to $1/2$ within its own future light cone and vanishes like a step function at its edges. 

To obtain the propagator in the massive case, one can expand the mass term perturbatively around the massless result. More specifically, rewriting (\ref{MassiveEquation}) as  $\Box G_m^R(t,x) = -\delta(t)\delta(x) + m^2 G_m^R(t,x)$, one can show that
\begin{equation}
G_m^R = G_0^R - m^2\, G_0^R * G_m^R\,,
\end{equation}
where $*$ denotes convolution in \M11. Iterating this relation gives
\begin{equation}
\label{massscattering}
G_m^R
=
G_0^R
- m^2\, G_0^R * G_0^R
+ m^4\, G_0^R * G_0^R * G_0^R
+ \cdots\,.
\end{equation}
In $1+1$ dimensions this expansion can be re-summed explicitly, yielding the well-known result
\begin{equation}
G_m^R(t,x)
=
\frac{1}{2}\,\theta(t)\,\theta(t^2-x^2)\,
J_0\!\left(m\sqrt{t^2-x^2}\right).
\end{equation}
Thus, the massive retarded propagator remains supported inside the future light cone but is no longer constant there.

To compute the discrete retarded propagator $\GGG_m^R$ on a causal set in an analogous manner, one may use the path-sum formalism introduced by Johnston \cite{johnston2008particle}, in which the massive propagator is constructed from a chosen massless propagator $\GGG_0^R$ by suitable reweighting of paths. In practice, however, one does not first solve a discrete analog of the inhomogeneous massless equation on the causal set itself. Rather, one identifies a suitable causal-set kernel whose sprinkling expectation value reproduces the known continuum massless retarded Green function in the relevant dimension and then constructs the massive propagator from it by iteration like in Eq. \eqref{massscattering}.

In the present case, comparing Eq.~\eqref{continuumprop} with Eq.~\eqref{Cdef}, it is therefore natural to choose the massless causal-set propagator as
\begin{equation}
\GGG_0^R=\frac{1}{2}\,\CCC,
\end{equation}
where $\CCC$ is the causal matrix. Although this choice is simple and successful, it also raises several conceptual questions upon closer inspection. Firstly, its motivation lies primarily in the fact that its sprinkling expectation reproduces the correct continuum behavior in $1+1$ dimensions,  but it is less clear why this quantity should be given in terms of $\CCC$, which is a non-local object. Secondly, one may ask why the matrix elements of $\GGG^R_0$ are taken to be \emph{exactly} equal to $1/2$ throughout the future light cone, rather than arising from a genuinely discrete construction whose values fluctuate locally and whose expectation approaches the continuum result only asymptotically.

In addition, a comparison across different spacetime dimensions raises further questions. In $3+1$ dimensions, the continuum propagator is supported only along the light cone and vanishes in its interior, naturally suggesting to express $\GGG^R_0$ in terms of $\LLL$~\cite{johnston2008particle}. However, in $2+1$ dimensions the continuum propagator decays algebraically with the proper time inside the light cone, which has led to more indirect prescriptions, for example, involving volume-related expressions built from $\CCC$ \cite{johnston2010quantum} or maximal-chain estimators of proper time based on $\LLL$~\cite{X_2017}. Thus, there is, at present, no single clearly preferred causal-set object from which the propagators in different dimensions arise uniformly. Rather, different dimensions seem to require different constructions. It is therefore natural to ask whether these prescriptions can be understood within a common underlying framework.

In this article, we take a first step toward placing these constructions on a more intrinsic
footing. We are guided by the view that the link matrix $\LLL$ in its role as the short-range adjacency matrix is more fundamental to the construction of the propagator than the long-range causal matrix $\CCC$. Indeed, $\LLL$ may be regarded as a discrete generator of causal propagation, since its powers encode paths of increasing length: $\LLL$ describes one-link connections, $\LLL^2$ two-link connections, and more generally $\LLL^k$ $k$-link connections. This makes $\LLL$ the more natural object from which to build physically relevant quantities whenever such a construction is possible.

Following these thoughts, we ask the question whether the massless propagators on a causal set are representable as a power series of the link matrix $\LLL$ irrespective of the embedding dimension. In the 3+1-dimensional case, this is trivially true, but the question arises whether such a representation is also possible in 2+1 and 1+1 dimensions. To do this, we would have to linearly combine the powers of $\LLL$ in such a manner that the correct asymptotic behavior is retrieved. As we show in the following, this idea can indeed be implemented for real scalar fields on Poisson sprinklings embedded in $1+1$ dimensional Minkowski spacetime.

The present work was prompted by a preliminary numerical study in which we investigated how the powers of the link matrix must be combined to approximate a propagator whose expectation value is asymptotically constant. We found that this is possible and that the resulting coefficients are very close to those of a simple exponential series. This led us to the proposal that the massless retarded propagator on a causal set embedded in \M11, up to normalization, is given by
\begin{equation}
\label{MainHypothesis}
\GGG^R_0 \propto \exp(\LLL)\,.
\end{equation}
Interestingly, a similar assumption recently appeared in a different context, namely in a causal-set approach to Algebraic Quantum Field Theory (AQFT)~\cite{buchanan2026causal}. There, the basic object is the past-link adjacency operator $B$, whose powers $B^n$ encode contributions from chains of length $n$, while the exponential $U(\lambda)=e^{\lambda B}=\sum_{n\ge 0}\frac{\lambda^n}{n!}B^n$ is introduced as a factorially damped causal transport operator on the Hasse diagram. In this sense, $B$ generates elementary one-link propagation, and the exponential naturally sums over longer causal paths with factorial weights. Although that construction is not intended as a derivation of the scalar retarded Green function considered here, it is conceptually very close to our proposal and provides independent support for the idea that exponentials of link-adjacency operators are natural objects in causal-set propagation.

We examine this proposal and the possible consequences in detail. The presentation is organized as follows. In Section II we review the combinatorial interpretation of powers of the link matrix and the associated expectation values for link-path counts. In Section III we develop the asymptotic analysis leading to the exponential link kernel in Eq. \eqref{MainHypothesis} and determine its normalization. Section IV presents supporting numerical evidence. In Section V we extend the construction to the massive case. Section VI discusses the inverse kernel and its possible interpretation as a discrete wave operator. Finally, we conclude with a summary and an outlook, including possible extensions to curved $1+1$ spacetimes. Additional mathematical details can be found in the Appendix.

\section{Length distribution of link Paths}

In a given causal set $\mathcal{C}$, a pair of causally related events $a \prec b$ can be connected by a multitude of different paths. The number of those paths from $a$ to $b$ which consist of exactly $k$ links equals the matrix element $(\LLL^k)_{ab}$ of the $k^{\text{th}}$ power of the link matrix $\LLL$. Obviously, the concatenation rule for this quantity is given by the matrix product
\begin{equation}
\label{LinkMatrixProduct}
(\LLL^{k+l})_{ab} \;=\; \sum_{c\in\langle a,b \rangle} (\LLL^k)_{ac}(\LLL^l)_{cb}\,,
\end{equation}
where the sum runs over all intermediate events contained in the causal diamond $\langle a,b \rangle$ spanned by $a$ and $b$.

Since sprinklings are statistically Lorentz-invariant, the average number of paths from $a$ to $b$ on a sprinkling embedded in a Minkowski space will depend only on the causal relativistic distance
\begin{equation}
\tau_{ab}\;=\;\sqrt{-x_\mu x^\mu} \;=\; \sqrt{(t_b-t_a)^2-(\vec x_b-\vec x_a)^2}\,.    
\end{equation}
This allows us to define the functions $F_k(\tau)$ as the expectation values of $(\LLL^k)_{ab}$ for a pair of points $a \prec b$ at distance $\tau>0$, that is
\begin{equation}
\label{FDef}
F_k(\tau) \;:=\; 
\Bigl\langle (\LLL^k)_{ab} \Bigr\rangle_{a \prec b\; \wedge\; \tau_{ab}=\tau}
\end{equation}
with $F_0(\tau)\equiv 0$. These functions will play an important role throughout this work. Note that  $F_k(\tau)$ is not a probability density, but the expectation value of the number of paths
consisting of exactly $k$ links. 

\begin{figure}
\includegraphics[width=100mm]{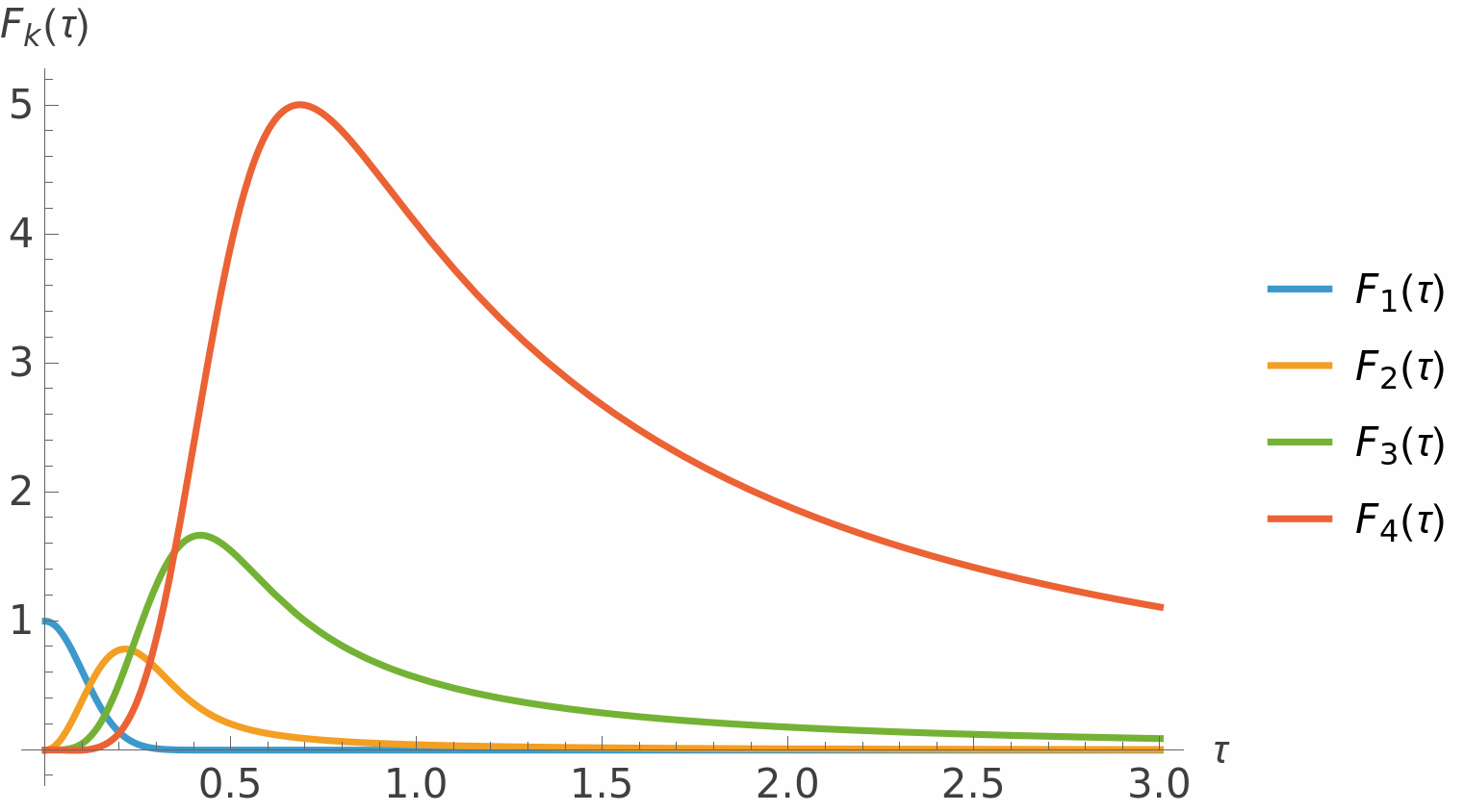}
\caption{The functions $F_1(\tau),\ldots,F_4(\tau)$ for $\rho=100$. 
Note that these functions are not normalized, which reminds us that they are not probability densities but rather expectation values for the number of paths.
\label{fig:F}
}
\end{figure}

Let us first consider $F_1(\tau)$. Since two events $a \prec b$ are connected by a single link if and only if there are no other events in the causal diamond $\langle a,b \rangle$ in between, and knowing that the sprinkled events are randomly distributed at constant density $\rho$, it is obvious that $F_1(\tau)$ is given by a Poisson distribution $F_1(\tau) = e^{-\rho V(\tau)}$. Here, $V(\tau)$ is the invariant volume of the causal diamond spanned by $a$ and $b$, which depends on the dimension of the embedding space. Specifically, in \M11, where $V(\tau)=\frac12\tau^2$, this leads to the known result
\begin{equation}
F_1(\tau) \;=\; e^{-\frac12 \rho\tau^2}\,.
\end{equation}
To compute the functions $F_2(\tau),F_3(\tau),\ldots$, we can use Eq.~(\ref{LinkMatrixProduct}), integrating over the position $(t_c,x_c)$ of the connecting intermediate point $c$. This leads us to the recursion relation
\begin{equation}
F_{k+l}(\tau_{ab})\;=\; \rho \int \d t_c \int \d x_c \, F_k(\tau_{ac}) F_l(\tau_{cb})\,,
\end{equation}
where the double integration runs over all points in the interior of the causal diamond $\langle a,b \rangle$. Changing variables $(t_c,x_c) \to (\tau_{ac},\tau_{cb})$ this recursion relation can also be written in bilinear form as
\begin{equation}
\label{BilinearProduct}
F_{k+l}(\tau)\;=\; \int_0^{\tau}\d \tau'\int_0^{\tau-\tau'}\d\tau''\, K(\tau;\,\,\tau',\tau'')\,\, F_k(\tau') \, F_l(\tau'')
\end{equation}
with the kernel
\begin{equation}
K(\tau;\,\,\tau',\tau'')\;=\;\frac{4\rho \,\tau'\tau''}{\sqrt{(\tau-\tau'-\tau'')(\tau-\tau'+\tau'')(\tau+\tau'-\tau'')(\tau+\tau'+\tau'')}}\,.
\end{equation}
Since this bilinear product plays an important role throughout this work, we will also use the compact notation
\begin{equation}
F_{k+l} \;=\; F_k \circledast F_l\,.
\end{equation}
Using this product, one can, for example, show that $F_2 = F_1 \circledast F_1$ is given by the exact expression
\begin{equation}
F_2(\tau) \;=\;
2 e^{-\frac{\rho  \tau ^2}{4}}\,
\text{Shi}\bigl(\tfrac{\rho \tau ^2}{4}\bigr)\,,
\end{equation}
where $\text{Shi}(z)=\int_0^z\tfrac{\sinh(t)}{t}\d t$ denotes the hyperbolic sine integral. Note that $F_1(\tau)$ is exponentially short-range while $F_2(\tau)$ already exhibits a long-ranged algebraic tail of the asymptotic form
\begin{equation}
F_2(\tau) \;\simeq\; \frac{4}{\rho\tau^2}
\quad \text{ for }
\tau \to\infty\,.
\end{equation}
For $k=3,4,\ldots$ one gets increasingly complex results. The first four functions $F_1(\tau),\ldots,F_4(\tau)$ are plotted in Fig.~\ref{fig:F}.
%

\section{Asymptotic behavior of $\exp(\LLL)$}
\label{Section:ExpL}
%
In \M11, the massless continuum propagator~\eqref{continuumprop} is constant and equal to $1/2$ in the interior of its future light cone, even for very large values of $\tau$. Attempting to construct a causal-set propagator exclusively in terms of the link matrix, it is therefore necessary to clarify whether it is in principle possible to reproduce this constancy as a statistical average by a suitable series of powers in $\LLL$. In what follows, we show that, within the asymptotic framework developed below, the requirement of asymptotic constancy selects the exponential series. Thus, the main result of
this work is the proposal that 
\begin{equation}
\GGG^R_0 \;\propto\; \exp(\LLL)\,.
\end{equation}
Moreover, we show that on a causal set, $\exp(\LLL)$ does not tend to $1/2$; instead, we find that it tends to $e^{-\gamma_E}\approx 0.561459$, where $\gamma_E$ denotes the Euler constant. This suggests normalizing the propagator by
\begin{equation}
\label{MainClaim}
\GGG^R_0 = \tfrac12 e^{\gamma_E} \exp(\LLL)\,.
\end{equation}
The most direct approach to establish (\ref{MainClaim}) would be to evaluate the power series $\exp(\LLL)=\sum_{k=0}^\infty \LLL^k/k!$ term by term. Since this power series translates directly into a corresponding series for functions $F_k(\tau)$ via Eq.~(\ref{FDef}), this would amount to showing that
\begin{equation}
\lim_{\tau\to\infty}\sum_{k=0}^\infty \frac{F_k(\tau)}{k!} \;=\; \text{const}\,.
\end{equation}
However, this would require us to calculate all $F_k(\tau)$ to all orders in $\tau$, which turned out to be not feasible in practice. Therefore, we use a different approach, which is motivated as follows. If we introduce a real parameter $\lambda>0$ and numerically investigate the asymptotic behavior of $\exp(\lambda \LLL)$, we find asymptotic power laws for large~$\tau$ with an exponent depending on $\lambda$. This suggests invoking the power law ansatz
\begin{equation}
\label{powerlawansatz}
G_\lambda(\tau)\;:=\;
\sum_{k=0}^\infty \frac{\lambda^k \, F_k(\tau)}{k!}
\;\simeq\;
\alpha(\lambda)\,\tau^{-\beta(\lambda)}
\quad \text{for }\tau\to\infty\,
\end{equation}
with an amplitude $\alpha(\lambda)$ and an exponent $\beta(\lambda)$, both continuously varying with $\lambda$. As we shall see in the following, this ansatz effectively performs a resummation of the large-$\tau$ behavior and allows analytical information to be extracted from the evolution equation in $\lambda$. More specifically, the idea of proving~(\ref{MainClaim}) amounts to establishing an evolution equation in the parameter $\lambda$ together with suitable initial conditions in $\lambda=0$. Showing that the ansatz~(\ref{powerlawansatz}) is self-consistent, this allows us to extract differential equations for $\alpha(\lambda)$ and $\beta(\lambda)$. Solving these differential equations, we finally determine $\alpha(\lambda)$ and $\beta(\lambda)$ at $\lambda=1$.
\paragraph{Evolution equation and initial conditions:}
Using Eq.~(\ref{BilinearProduct}), one can easily show that the function $G_\lambda(\tau)$ obeys the linear differential equation
\begin{align}
\label{pde}
\frac{\partial G_\lambda(\tau)}{\partial \lambda}
\;&=\; \nonumber
\sum_{k=1}^{\infty}\frac{k \lambda^{k-1} F_k(\tau)}{k!}
\;=\;\Bigl( F_1 \circledast \sum_{k=1}^{\infty}\frac{ \lambda^{k-1} F_{k-1}(\tau)}{(k-1)!} \Bigr)(\tau)
\\ \;&=\;
\Bigl( F_1 \circledast \sum _{k=0}^{\infty}\frac{ \lambda^{k} F_{k}(\tau)}{k!} \Bigr)(\tau)
\;=\; \Bigl( F_1 \circledast G_\lambda(\tau) \Bigr)(\tau)\,.
\end{align}
For small $\lambda$, the series~(\ref{powerlawansatz}) is approximately given by $G_\lambda(\tau) = F_0(\tau) + \lambda F_1(\tau) + \frac{\lambda^2}{2}F_2(\tau) + \orderof{\lambda^3}$. Here, $F_0$ and $F_1$ are both short-range in $\tau$, meaning that the lowest-order term that generates a long-range contribution is
\begin{equation}
F_2(\tau) \;\sim \; \frac{4}{ \rho\tau^2} \qquad \text{for } \tau\to\infty\,.
\end{equation}
From this, we conclude that for $\lambda\to 0$ the solution should behave as
\begin{equation}
\label{initialbehavior}
G_\lambda(\tau) \;\sim \; \frac{2\lambda^2}{ \rho\tau^2}\qquad \text{for } \tau\to\infty\,.
\end{equation}
This defines the initial condition of the partial differential equation (\ref{pde}) at $\lambda=0$ for large values of $\tau$.
\paragraph{Power law ansatz:}
%
In order to solve Eq.~(\ref{pde}), we now assume that $G_\lambda(\tau)\;\simeq\; \alpha(\lambda) \, \tau^{-\beta(\lambda)}$ for large $\tau$ with two functions $\alpha(\lambda)$ and $\beta(\lambda)$ to be determined. The calculated behavior for small $\lambda$ in~(\ref{initialbehavior}) implies that the initial value of the exponent $\beta(\lambda)$ is given by
\begin{equation}
\beta(0)=2\,.
\end{equation}
Furthermore, Eq.~(\ref{initialbehavior}) tells us that the amplitude $\alpha(\lambda)$ is initially zero and then begins to increase as
\begin{equation}
\label{quadratic}
\alpha(\lambda) \simeq \frac{2\lambda^2}{\rho} \qquad \text{for } \lambda \to 0 \,.
\end{equation}
The power law ansatz~(\ref{powerlawansatz}) implies that the derivative of $G_\lambda(\tau)$ on the left hand side of Eq.~(\ref{pde}) is given by
\begin{equation}
\label{derivative}
\frac{\partial G_\lambda(\tau)}{\partial \lambda}\;=\;
\Bigl( \alpha'(\lambda ) -\alpha (\lambda )\beta'(\lambda )
   \ln (\tau )  \Bigr) \, \tau ^{-\beta (\lambda )}
   \qquad \text{ for } \tau\to\infty\,.
\end{equation}
However, the ansatz~(\ref{powerlawansatz}) is self-consistent if and only if the integral $F_1 \circledast G_\lambda$ on the right-hand side of Eq.~(\ref{pde}) also produces two additive contributions with exactly the same structure, allowing the differential equations for $\alpha(\lambda)$ and $\beta(\lambda)$ to be extracted by comparing coefficients. For this, $F_1 \circledast G_\lambda$ has to be calculated in the limit case of large $\tau$. In fact, as shown in Appendix~\ref{AppendixCalculation} by a series of suitable approximations, the right-hand side of Eq.~(\ref{pde}) evaluated for large $\tau$ exhibits the asymptotic behavior
\begin{equation}
\label{claim1}
\bigl(G_\lambda \circledast F_1\bigr)(\tau)
\;=\;\bigl(F_1  \circledast G_\lambda\bigr)(\tau)
\;=\;G_\lambda(\tau)\,\Bigl( 2\ln\tau+C_\lambda \Bigr)\,,
\end{equation}
where
\begin{equation}
\label{claim2}
C_\lambda \;=\; \ln \rho+\ln 2 - \gamma_E
+ \psi\Bigl(\frac{3-\beta(\lambda)}{2}\Bigr)
- \psi\Bigl(\frac{2-\beta(\lambda)}{2}\Bigr)
- 2\psi\Bigl(2-\beta(\lambda)\Bigr)
\end{equation}
is a $\tau$-independent constant, in which $\gamma_E$ is the Euler constant and $\psi(z)=\Gamma'(z)/\Gamma(z)$ denotes the digamma function. 
\paragraph{Solving the differential equations for $\alpha(\lambda)$ and $\beta(\lambda)$:}
%
Comparing Eqs.~\eqref{derivative} and \eqref{claim1} in the limit of large $\tau$, the comparison of the logarithmic contributions leads to the differential equation $\beta'(\lambda)=-2$. Since Eq.~(\ref{initialbehavior}) tells us that $\beta(0)=2$, we can conclude that
\begin{equation}
\label{solution1}
\beta(\lambda) \;=\; 2-2\lambda \;.
\end{equation}
In particular, setting $\lambda=1$ we obtain $\beta(1)=0$, proving that $G_1(\tau)$ and therewith $\langle\exp(\LLL)\rangle$ are asymptotically constant in the limit $\tau\to \infty$.

Comparing the coefficients of the remaining constant contributions leads to the second differential equation
\begin{equation}
\frac{\alpha '(\lambda )}{\alpha(\lambda)}
\;=\; C(\lambda)
\end{equation}
where the r.h.s., after inserting (\ref{solution1}), is now given by
$ C_\lambda = \log\rho
+\log 2
-\gamma_E
+\psi\bigl(\lambda+\tfrac12\bigr)
-\psi\bigl(\lambda\bigr)
-2\psi\bigl(2\lambda\bigr)$.
The solution of this differential equation reads
\begin{equation}
\label{alphaAmplitude}
\alpha(\lambda)\;=\; A\frac{(\rho/2)^{\lambda-1}\,e^{\gamma_E(1-\lambda)}}{\Gamma^2(\lambda)}\,.
\end{equation}
To determine the integration constant $A$, we expand this expression to small values of $\lambda$, obtaining
\begin{equation}
\alpha(\lambda)\;\approx\; \frac{2A e^{\gamma_E}\lambda^2}{\rho} + \orderof{\lambda^3}\,.
\end{equation}
On the other hand, Eq.~(\ref{quadratic}) tells us that $\alpha(\lambda)$ increases for small $\lambda$ as $ \frac{2\lambda^2}{\rho}$, which implies that $A=e^{-\gamma_E}$, leading us to the final solution
\begin{equation}
\alpha(\lambda)\;=\; \frac{e^{-\lambda \gamma_E}\,\,(\rho/2)^{\lambda-1}}{\Gamma^2(\lambda)}\,.
\end{equation}
with the special $\rho$-independent value
\begin{equation}
\label{alpha}
\alpha(1) \;=\; e^{-\gamma_E} \;\approx\; 0.561459\,.
\end{equation}
Thus, we have shown that $G_1(\tau)$ tends to the constant value $e^{-\gamma_E}$ in the limit $\tau\to\infty$, suggesting that the retarded propagator on the discrete causal set expressed in terms of the link matrix $\LLL$ is given by 
\begin{equation}
\label{MainResult}
\GGG^R_0 \;=\; \tfrac{1}{2} e^{\gamma_E}\exp(\LLL)\,.
\end{equation}
\paragraph{Consistency check:}
%
As an additional check of the results obtained above, we expand both sides of Eq.~(\ref{claim1}) as power series in $\lambda$. By matching the coefficients on both sides, we can recursively compute the exact expressions for $F_3(\tau),F_4(\tau),\ldots$ with the help of an algebraic computer system, which produces increasingly complex expressions. Nevertheless, these explicit expressions allow us to extract the behavior for large $\tau$ for $k=2,\ldots,7$, leading to the conclusion that
\begin{equation}
F_k(\tau) \;\simeq\; k(k-1)\,\frac{\bigl[ \ln V(\tau) \bigr]^{k-2}}{V(\tau)} \qquad \text{for } \tau\to\infty\,,
\end{equation}
where $V(\tau)=\frac12\rho\tau^2$ is again the volume of the causal diamond in \M11. Consequently, the asymptotic behavior of $G_\lambda(\tau)=\sum_{k=0}^\infty \frac{\lambda^k}{k!}F_k(\tau)$ is given by
\begin{equation}
G_\lambda(\tau) \sim \lambda^2 [V(\tau)]^{\lambda-1} \qquad \text{for } \tau\to\infty\,,
\end{equation}
consistently reproducing our previous finding that the function is asymptotically constant for $\lambda=1$.
%
%
\section{Supporting numerical evidence}
%
\begin{figure}
\includegraphics[width=75mm]{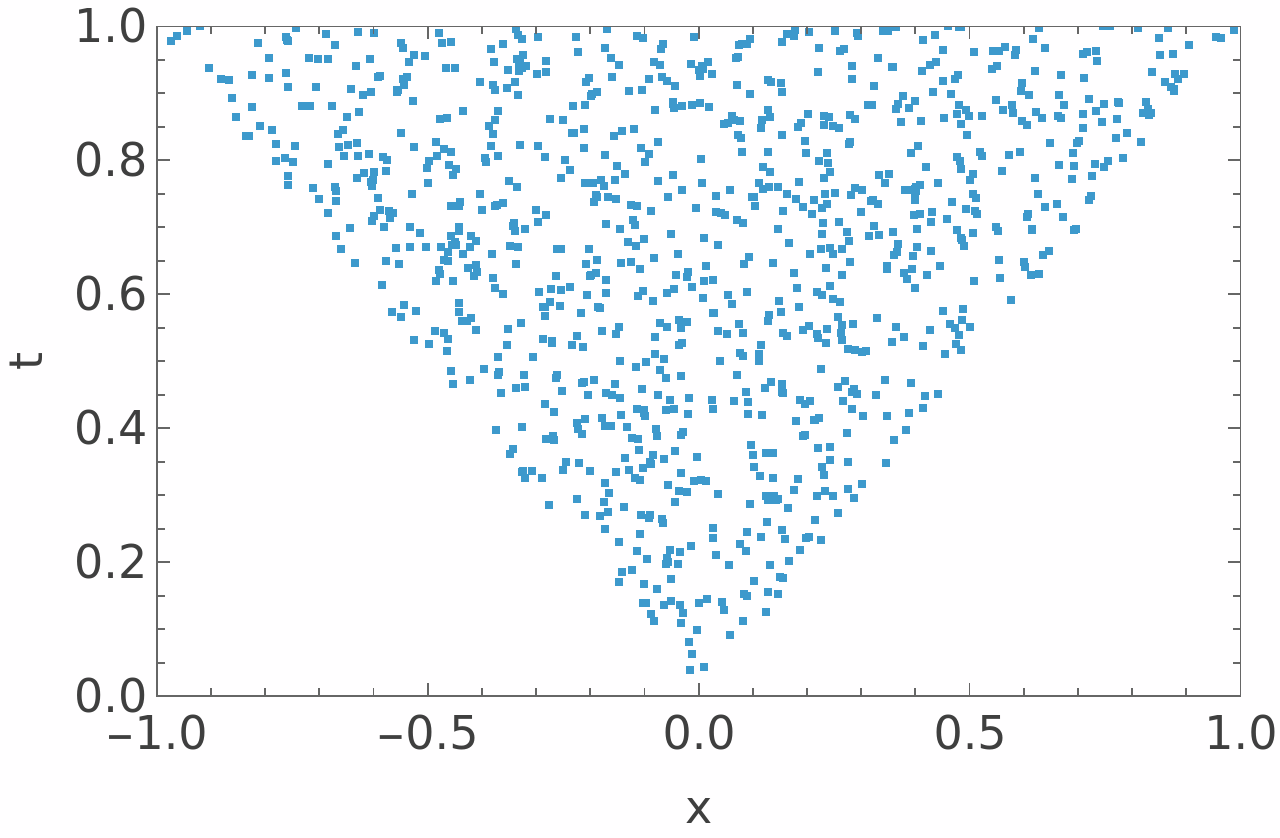}
\caption{Sprinkling with density $\rho=1024$ generated in a triangular region embedded in \M11.
\label{fig:sprinkling}
}
\end{figure}
%
%
To verify the results obtained above, we analyze $\exp(\LLL)$ numerically. To do this, we generate sprinklings in a triangular region with unit volume in $(-1,1)\times(0,1) \subset \mathbb M^{1+1}$, in which we randomly distribute $N=\rho V$ events at constant density $\rho$. Fig.~\ref{fig:sprinkling} shows such a sprinkling with $N=1024$ events. The generated events are then sorted by ascending time coordinates to ensure that the causal matrix $\CCC$ of size $N\times N$ is strictly upper-triangular. 

Having determined the causal matrix $\CCC$, we compute the following matrices of dimension $N \times N$:
\begin{itemize}
\item 
First, we calculate the link matrix~$\LLL$ according to Eq.~(\ref{Ldef}).
\item 
With $\LLL$ at hand, we then calculate the matrix $\KKK$ by
\begin{equation}
\KKK_{ij} \;=\;  \max \{ k \in \mathbb{N} \mid (\LLL^k)_{ij} \neq 0 \}\,.
\end{equation}
Here, the matrix element $\KKK_{ij}$ is the maximum number of links between the two events $i$ and $j$, which serves as a discrete proper time measure \cite{rideout2010dynamics}.
\item 
For comparison, we define the matrix $\SSS$ by
\begin{equation}
\SSS_{ij} \;=\;
\begin{cases}
\sqrt{(t_j-t_i)^2-(x_j-x_i)^2} & \text{if } \CCC_{ij}=1 \\
0 & \text{otherwise}
\end{cases}
\end{equation}
which holds the corresponding relativistic distance between the events $c_i$ and $c_j$ in the embedding space of the sprinkling.
\item
Finally, we compute $\exp(\lambda\LLL)$ numerically. To this end, we use the \texttt{Eigen} C++ template library~\cite{eigen} with hardware acceleration and multi-core threading on a workstation. 
\end{itemize}
Having calculated these matrices, we then test whether the sprinkling average $\langle \exp(\LLL)_{ij} \rangle$, viewed as a function of the embedding proper time $\tau_{ij}$, becomes approximately constant and
approaches $e^{-\gamma_E}$. This can be done by creating a histogram as follows. Since the relativistic distance $\tau_{ij}$ within the triangular sprinkling is restricted to the interval $[0,1]$, we first divide this interval into 500 equidistant bins and assign the matrix elements $\tau_{ij}$ to these bins. Then the corresponding matrix elements $\exp(\LLL)_{ij}$ are averaged within each of the bins, allowing us to visualize the expectation value of $\exp(\LLL)$ as a function of $\tau$. To suppress fluctuations, the results are additionally averaged across $M$ statistically independent sprinklings.
%
\begin{figure}
\includegraphics[width=76mm]{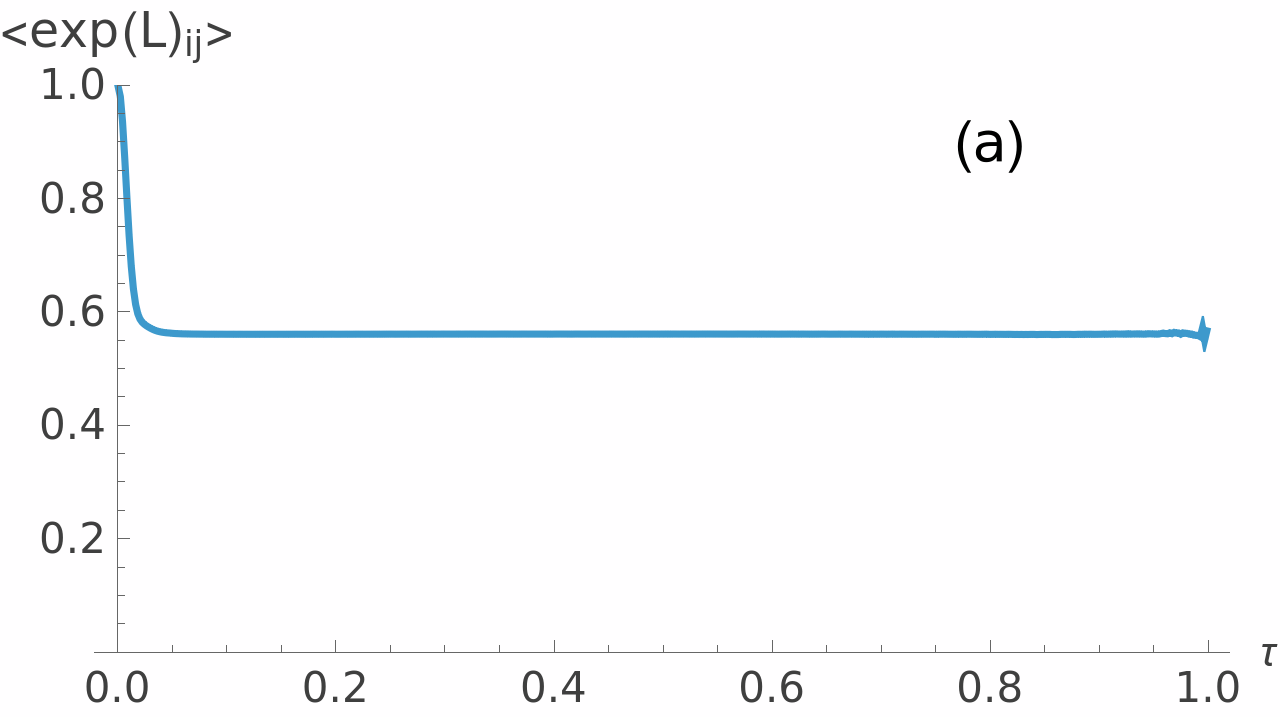}
\includegraphics[width=98mm]{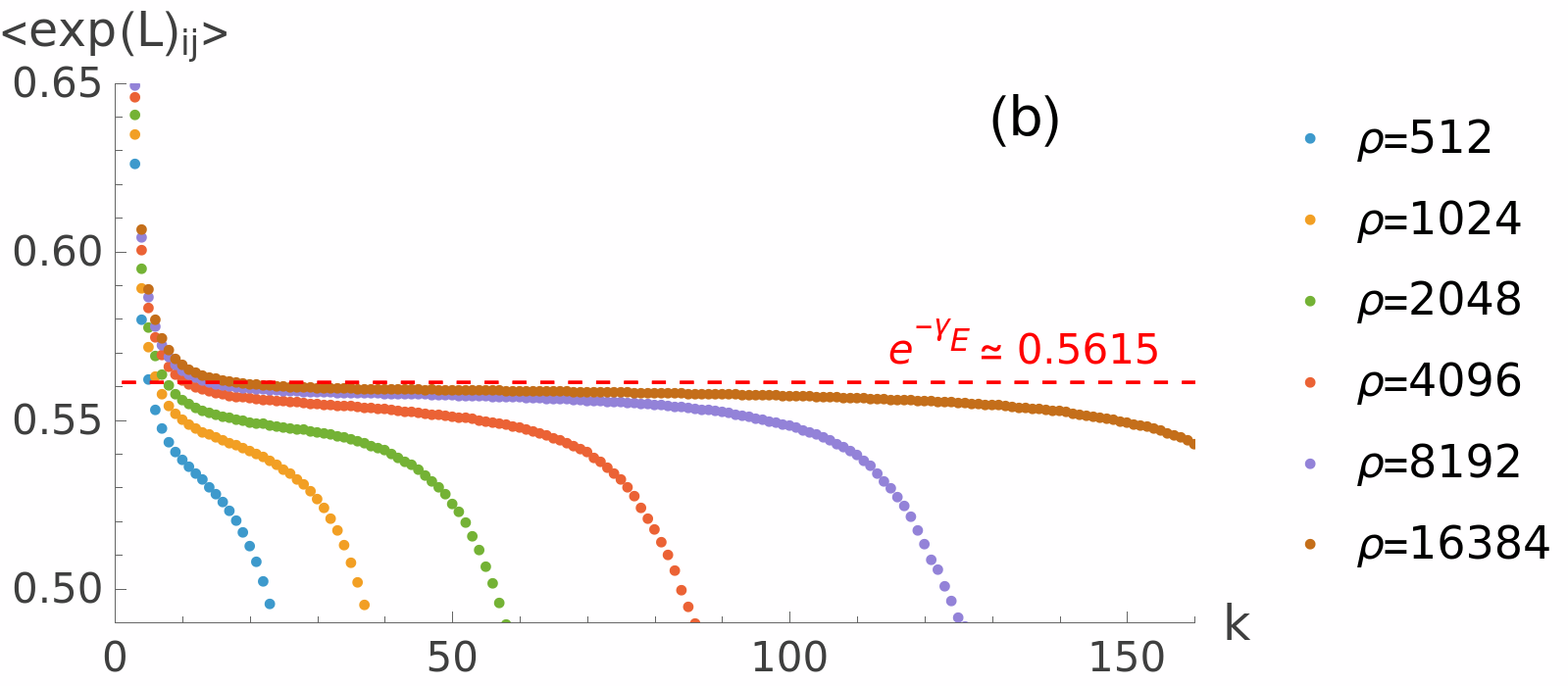}
\caption{Numerical confirmation of asymptotic constancy. (a) Expectation value of $\exp(\LLL)$ as a function of the relativistic distance $\tau$ in the embedding space \M11 averaged over $1000$ sprinklings with $N=16384$ events. (b) Expectation value of $\exp(\LLL)$ as a function of the maximal number of links $k=\KKK_{ij}$ for different sprinkling densities $\rho$. 
\label{fig:expl}
}
\end{figure}

Our results are shown in Fig.~\ref{fig:expl}a, where we calculated $\langle \exp(\LLL) \rangle_{ij}$ on a sprinkling with $N=16384$ events averaged over $M=1000$ runs. As can be seen, this quantity quickly decreases from $1$ to an approximately constant value close to $e^{-\gamma_E}\approx 0.561459$, confirming the calculation presented in the preceding Section.

An aspect of this result that remains conceptually unsatisfactory is that we have expressed the expectation value $\langle e^\LLL_{ij} \rangle$ as a function of the relativistic distance $\tau$ in the embedding space. According to the philosophy of causal networks, it is preferable to express all results without resorting to coordinates and distances in the embedding space. To do this, the embedding distance $\tau_{ij}$ must be replaced by its discrete analog, the maximum number of links $k=\KKK_{ij}$. Since the main findings of this article apply to the limit of large distances $\tau$, and since it is also known that $k$ and $\tau$ are proportional to each other, $\langle \exp(\LLL)_{ij} \rangle$ will inevitably exhibit the same asymptotic behavior as a function of $k$, tending towards the constant $e^{-\gamma_E}$. 

To verify this expectation numerically, we have plotted $\langle e^\LLL_{ij} \rangle$ versus $k$ in Fig.~\ref{fig:expl}b. As can be seen, the expectation value tends indeed to the calculated constant for sufficiently large densities $\rho$. However, to see this beyond a doubt, one must go to extremely high densities, that is, the deviations are more pronounced here than when plotting the data against $\tau$. These significantly larger deviations for finite $\rho$ may be attributed to non-trivial correlations between the matrix elements of $\exp(\LLL)_{ij}$ and $\KKK_{ij}$ and require further investigation. 
%
\section{Massive case}
We now turn to the massive scalar fields on Poisson sprinklings embedded in \M11. As outlined in the Introduction, the massive retarded propagator in the continuum is obtained from the massless one by repeated mass scattering. The causal set version proceeds in complete analogy, except that continuum convolutions must be replaced by sums over sprinkled elements. This amounts to replacing integrals by sums according to
\begin{equation}
\int d^2 z \;\longrightarrow\; \frac{1}{\rho}\sum_z \,.
\end{equation}
Thus, the continuum mass expansion~\eqref{massscattering} translates directly into the causal set expression
\begin{equation}
\GGG_m^R
\;=\;
\sum_{n=0}^{\infty}
\left(-\frac{m^2}{\rho}\right)^n
\left(\KKK_0^R\right)^{n+1},
\end{equation}
where matrix multiplication plays the role of discrete convolution. This is again a geometric series and may therefore be written formally as
\begin{equation}
\GGG_m^R
\;=\;
\GGG_0^R
\left(
\id+\frac{m^2}{\rho}\GGG_0^R
\right)^{-1}.
\end{equation}
Hence, the massive causal set propagator is obtained from the massless one by exactly the same scattering prescription as in the continuum, with the only modification being the factor $m^2/\rho$, which reflects the replacement of spacetime integrals by sums over sprinkled points.
\begin{figure}[t!]
\includegraphics[width=80mm]{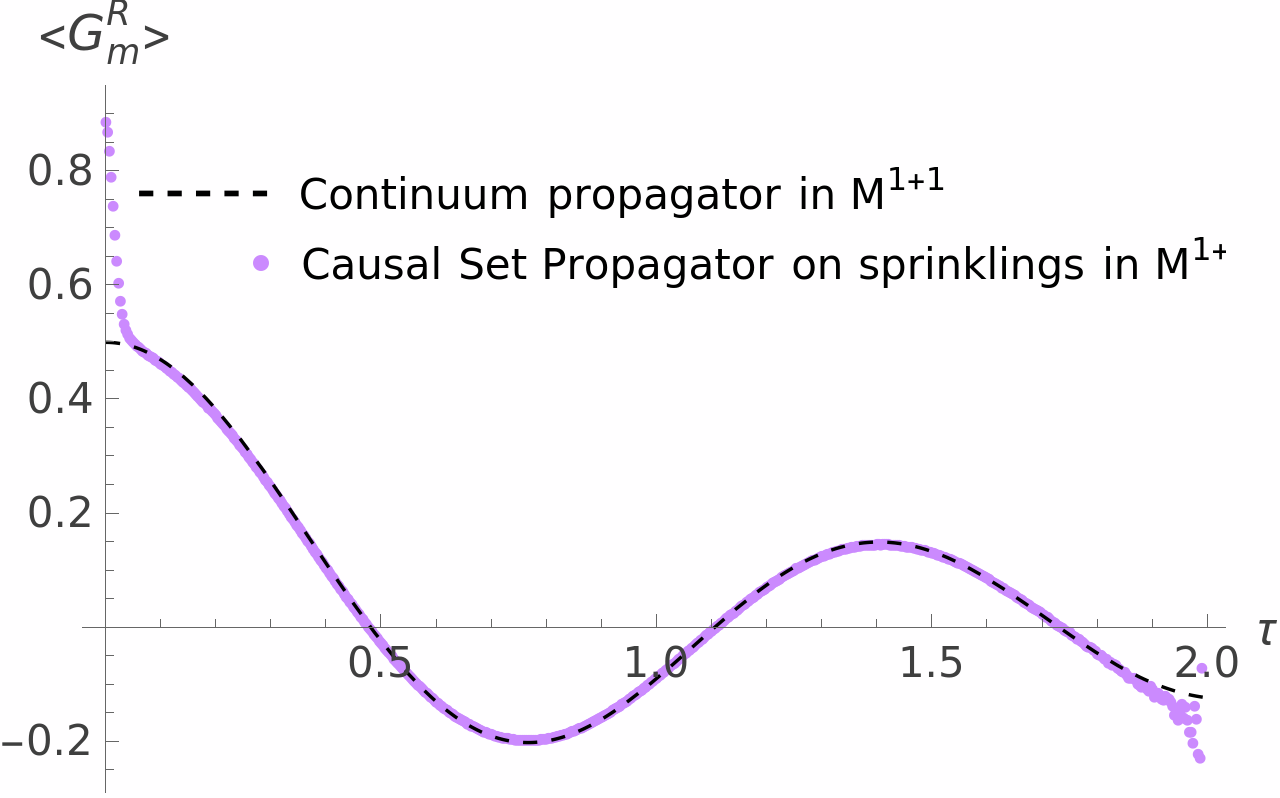}
\caption{Average of the massive retarded causal set propagator $\langle \GGG_m^R\rangle$ plotted as a function of the proper time $\tau$, together with the corresponding continuum retarded propagator in \M11. The causal set data were obtained from Poisson sprinklings with density $\rho=5000$, using mass $m=5$, $600$ bins, and averaging over $30$ realizations.
\label{fig:masspropavg}}
\end{figure}
Inserting the massless propagator obtained in the previous section, given by
\begin{equation}
\GGG_0^R = a\,\exp(\LLL),
\qquad
a=\frac12 e^{\gamma_E}.
\notag
\end{equation}
it follows that the massive propagator takes the form
\begin{equation}
\GGG_m^R
=
a\,\exp(\LLL)
\left(
\id+\frac{a m^2}{\rho}\exp(\LLL)
\right)^{-1}.
\end{equation}
This construction is structurally analogous to Johnston’s retarded causal-set propagator for
nonzero mass \cite{johnston2010quantum},  the difference being that the massless kernel is now taken to be the normalized exponential of the link matrix, so the elementary hop constant becomes
\begin{equation}
a=\frac12 e^{\gamma_E}.
\end{equation}
Using sprinkling density $\rho=5000$, mass $m=5$, $600$ bins, and averaging over $30$ independent runs, we show numerically that the averaged causal set massive propagator $\langle \GGG_m^R\rangle$ is in very good agreement with the continuum retarded propagator (see Fig. \ref{fig:masspropavg}). The visible deviation between them in the beginning and the end is naturally explained by the relatively small number of sprinkled events in these regions.

\section{Discretized Equation of Motion}
%
With the results obtained so far, we have shown that the retarded continuum propagator of a massless scalar field, which is constant within its light cone, can be approximated on a causal set by the exponential function of the link matrix together with a suitable normalization factor. However, at this point, it is not yet clear whether $\exp(\LLL)$ should be interpreted as a physically meaningful propagator in the full dynamical sense. To strengthen that interpretation, one would like to identify a corresponding discrete equation of motion for which the proposed kernel acts as an inverse in an appropriate sense. In this Section, we shall discuss the first steps in this direction.

In the continuum case, the massless retarded propagator $G_0^R(t,x)$ in \M11 is a solution of the inhomogeneous wave equation $\quabla G_0^R(t,x)=-\delta(t) \delta(x)$, which means that the propagator is essentially the inverse of the differential operator~$\quabla$. Thus, if $\GGG_0^R$  is indeed represented by a normalized exponential of $\LLL$,  the associated inverse
kernel naturally suggests a candidate discrete analog $\DDD$ of the continuum wave operator,
namely~$\DDD \propto \exp(-\LLL)$. Moreover, since the propagator in~(\ref{MainResult}) was normalized by $\GGG^R_0 = \tfrac{1}{2} e^{\gamma_E}\exp{\LLL}$, it is natural to expect that the matrix $\DDD$ comes with the reciprocal normalization constant, that is,
\begin{equation}
\label{MainClaim2}
\DDD \;=\; 2e^{-\gamma_E} \exp(-\LLL)\,.
\end{equation}
To determine whether this proposal is meaningful, we must verify two points. First, $\DDD$ should be short-ranged in $\tau$. Second, in the continuum limit of large sprinkling density and averaging over many realizations, it should reproduce the expected behavior of the continuum wave operator $\quabla=\partial_x^2-\partial_t^2$. 

\begin{figure}
\includegraphics[width=80mm]{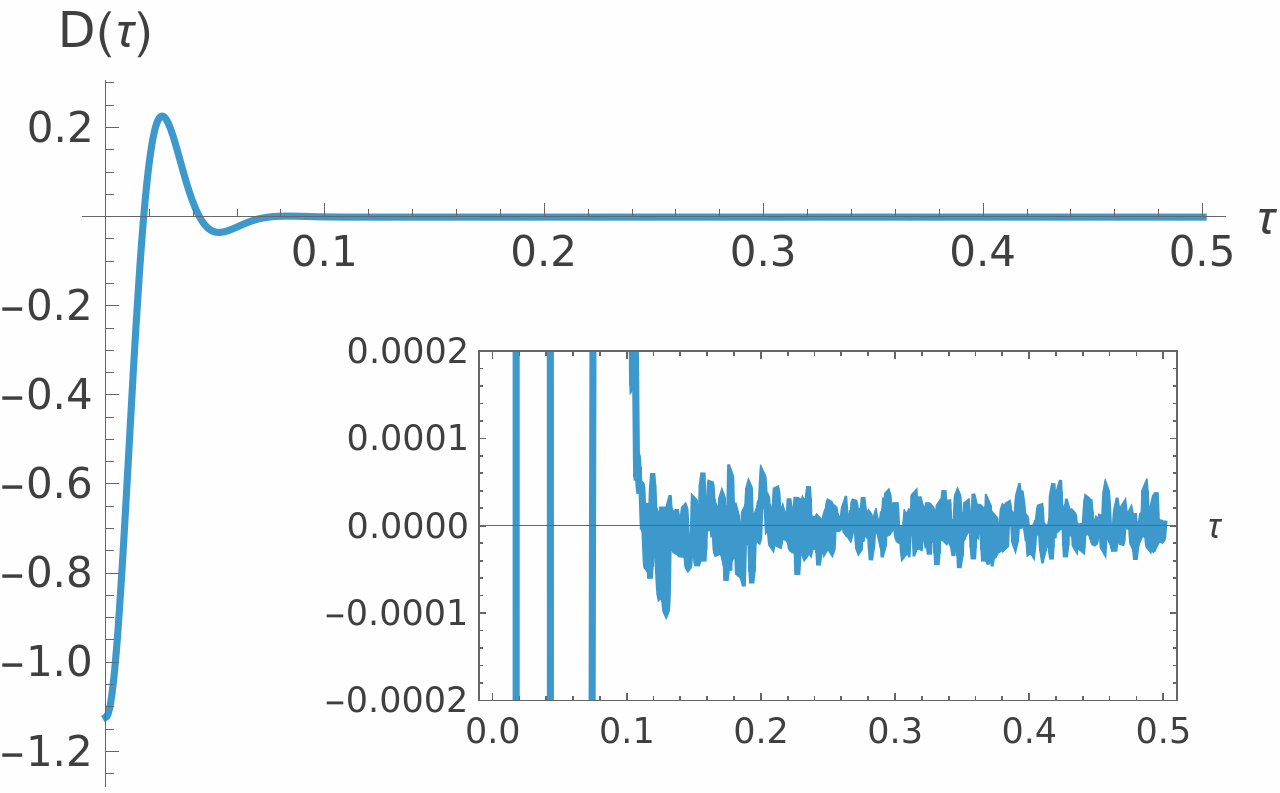}
\caption{Expectation value $D(\tau)=\langle 2 e^{-\gamma_E}\exp(-\LLL) \rangle$ plotted as a function of the relativistic distance $\tau$ in the embedding space of a sprinkling with 8192 events averaged over 2400 independent sprinklings. To illustrate the accuracy, the inset shows a magnification of the same data.
\label{fig:quabla}}
\end{figure}

To demonstrate that $\DDD$ is, in fact, a short-range operator, we define the expectation value 
\begin{equation}
\label{DDef}
D(\tau) \;:=\; 2e^{-\gamma_E}\,
\Bigl\langle \bigl(\exp(-\LLL)\bigr)_{ab} \Bigr\rangle_{a \prec b\, \wedge\,\tau_{ab}=\tau} 
\end{equation}
as a function of $\tau$, following exactly the same procedure as for $G_0^R(\tau)$ in the preceding Section. As shown in Fig.~\ref{fig:quabla}, this function first oscillates and then quickly tends to zero. This short-range characteristics can also be understood analytically if we use the results of Section \ref{Section:ExpL}, the only difference being that we now set $\lambda=-1$. Then, the divergent gamma function in~(\ref{alphaAmplitude}) implies that $\alpha(-1)=0$, which means that the amplitude of the long-range component of $\langle e^{-\LLL} \rangle$ vanishes as $\tau\to\infty$. This confirms that $\DDD$ is indeed an exponentially decaying short-range operator on the causal set.

\begin{figure}
\includegraphics[width=90mm]{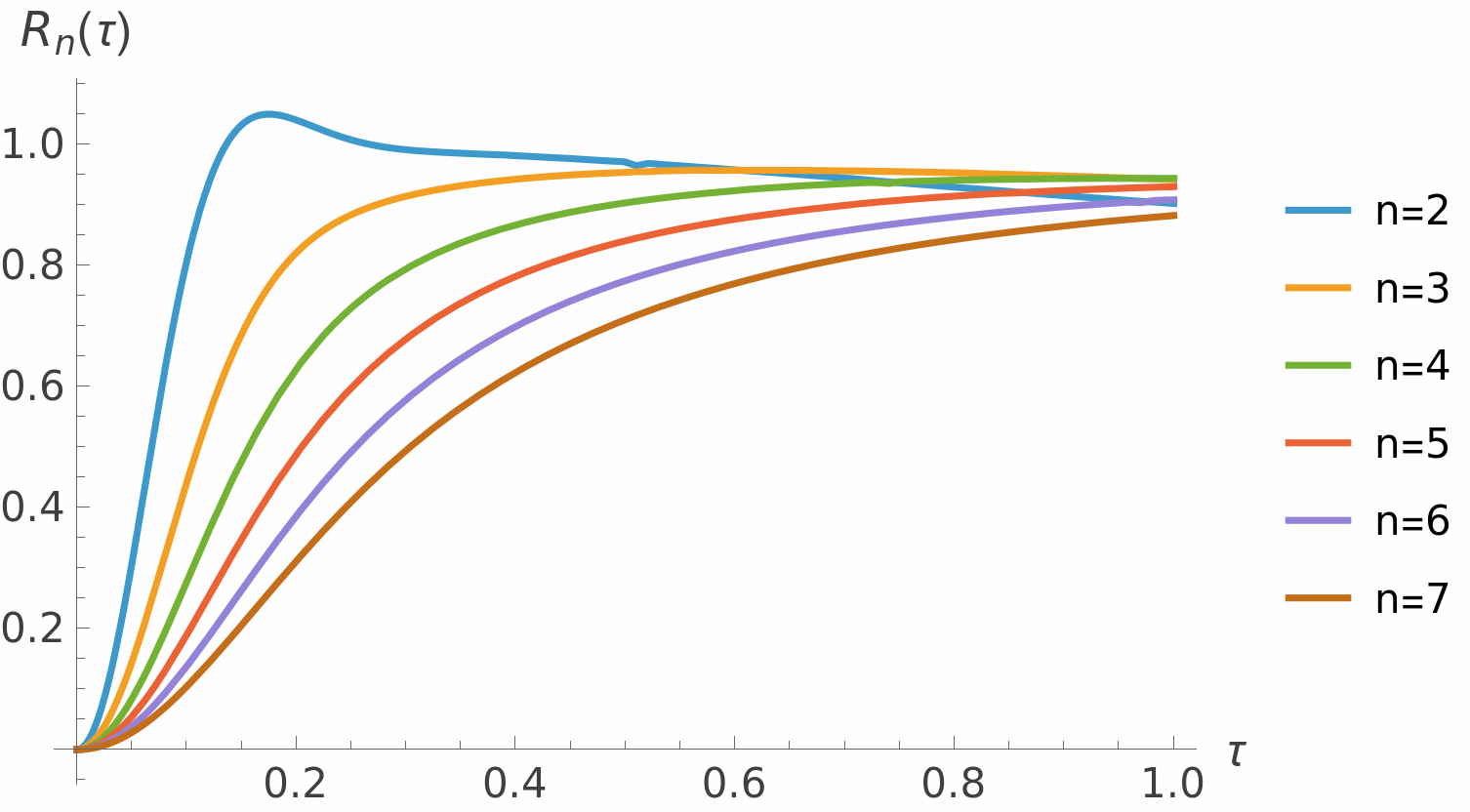}
\caption{Numerical check of $\DDD$ acting on a power $\tau^n$ for $n=2,3,\ldots,7$, divided by the expected continuum result. As can be seen, the results are consistent with the assumption that the quotients tend to 1, confirming that $\DDD$ approximates $\quabla$ on $\tau$-dependent functions.
\label{fig:dalembert}}
\end{figure}

In order to show that $\DDD$ indeed approximates the continuum differential operator $\quabla$, recall that in polar coordinates $t=\tau\cosh\eta$ and $x=\tau\sinh\eta$, the d'Alembert operator is given by
\begin{equation}
\label{DefinitionQuabla}
\quabla 
\;=\; \frac{\partial^2}{\partial x^2}-\frac{\partial^2}{\partial t^2}
\;=\; -\frac{\partial^2}{\partial\tau^2} - \frac{1}{\tau}\frac{\partial}{\partial \tau}+\frac{1}{\tau^2}\frac{\partial^2}{\partial\eta^2}\,.
\end{equation}
Restricting our analysis to test functions that depend exclusively on powers of $\tau$, we would like to verify whether the continuum identity
\begin{equation}
\quabla \tau^n  \;=\;  -n^2 \tau^{n-2} \, , \quad n=2,3,\ldots
\end{equation}
in the continuum can be faithfully reproduced with $\DDD$ acting on $\tau^n$ on a causal set in the continuum limit. To this end, we have to show that the bilinear product $D(\tau) \circledast \tau^n$ approaches $n^2 \tau^{n-2}$ for large $\tau$. However, we must take into account that $e^{-\LLL}$ includes the identity to the leading order, which is not accounted for in $D(\tau)$. This means that we have to add this contribution by hand. Thus, we have to show that
\begin{equation}
\label{RelationToShow}
D(\tau) \circledast \tau^n  +
2 e^{-\gamma_E} \tau^n  
\;\simeq\; n^2 \tau^{n-2} \qquad \text{ for } \tau \to\infty\,,
\end{equation}
where
\begin{equation}
D(\tau) \circledast \tau^n \;=\;
\int_0^\tau \d\tau'\,D(\tau')  
\underbrace{\int_0^{\tau-\tau'}\d\tau''   \,K(\tau;\, \tau',\tau'')\,[\tau'']^n}_{=: \; I_n(\tau,\tau')}\,.
\end{equation}
The inner integration can be carried out, giving
\begin{equation}
I_n(\tau,\tau') \;=\;
\frac{\sqrt{\pi } \, n \,\rho}{\tau '+\tau }\, \Gamma \bigl(\tfrac{n}{2}\bigr)
 \,  \tau'\, \left(\tau -\tau '\right)^{n+1} \,
   _2\tilde{F}_1\Bigl(\tfrac{1}{2},\tfrac{n+2}{2};\tfrac{n+3}{2}
   ;\tfrac{\left(\tau -\tau '\right)^2}{\left(\tau +\tau
   '\right)^2}\Bigr)
\end{equation}
with $_2\tilde{F}_1$ denoting the regularized hypergeometric series. Carrying out the outer integration numerically on a sprinkling with 512 events averaged over a large number of $10^6$ independent realizations, we plotted the ratios
\begin{equation}
R_n(\tau) \;=\;
\frac{2 e^{-\gamma_E} \tau^n + 
\bigl(D \circledast \tau^n\bigr)_{(\tau)} }
{n^2 \tau^{n-2}}
\end{equation}
versus $\tau$ for $n=2,3,\ldots,6$ in Fig.~\ref{fig:dalembert}. If~\eqref{RelationToShow} applies, then these ratios should tend to $1$ as $\tau\to\infty$. Our numerical results are consistent with this assertion, supporting the claim that $\DDD=2e^{-\gamma_E}\exp(-\LLL)$ approximates the continuum wave operator on this class of $\tau$-dependent test functions. The observed numerical deviations can be attributed to the finiteness of the sprinkling.

\begin{figure}[b]
\includegraphics[width=100mm]{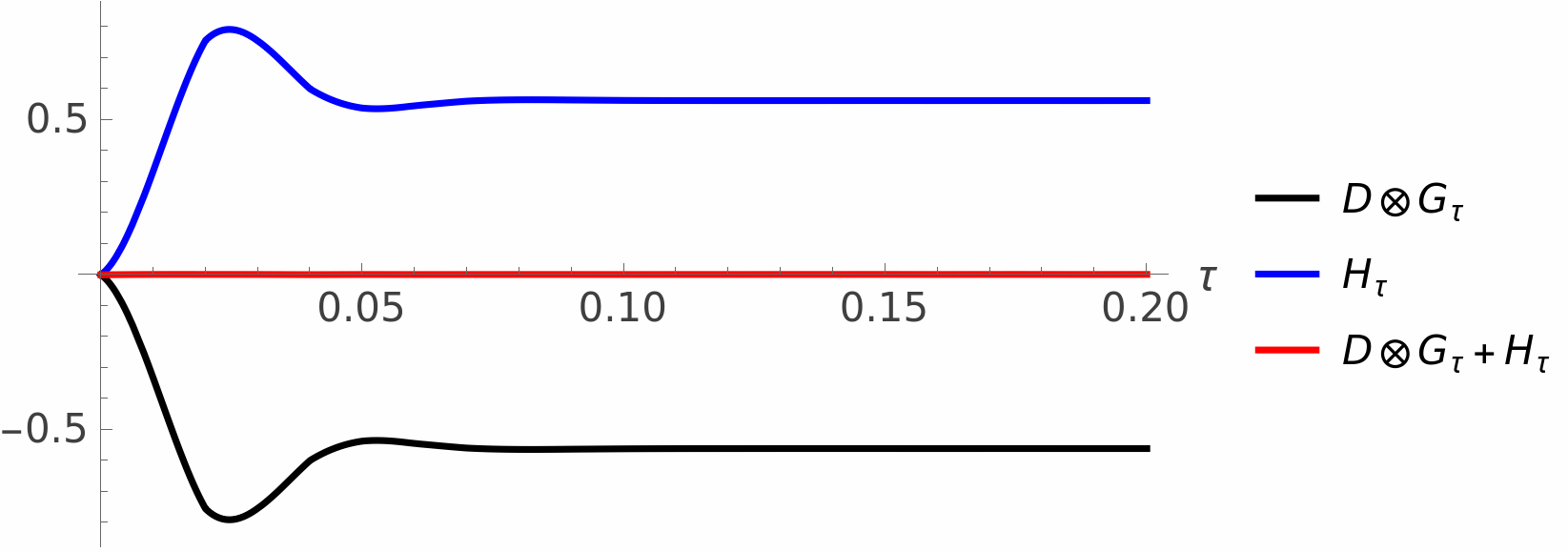}
\caption{Numerical verification on the equation of motion~(\ref{eom}) on a sprinkling with 8192 events in \M11. As can be seen, the two terms cancel one another.
\label{fig:diffeq}}
\end{figure}

To further support our claim that $\DDD$ approximates $\quabla$, we verify whether the inhomogeneous wave equation $\quabla G_0^R(t,x)=-\delta(t) \delta(x)$ in the continuum can be recovered in terms of the functions $D(\tau)$ and $G_0(\tau)$ on the causal set. Once again we have to take into account that these functions represent only the off-diagonal matrix elements for $\tau>0$, that is, including the normalization we have the correspondence
\begin{equation}
\begin{split}
G_0^R(\tau) \quad & \leftrightarrow \quad \tfrac12 e^{\gamma_E} \bigl( \exp(\LLL)-\id \bigr) \\
D(\tau) \quad & \leftrightarrow \quad  2 e^{-\gamma_E} \bigl( \exp(-\LLL)-\id \bigr) \,.
\end{split}
\end{equation}
Hence the discrete counterpart of the massless inhomogeneous wave equation reads
\begin{equation}
\label{eom}
\bigl(D \circledast G_0^R\bigr)_{(\tau)} + H(\tau) \;=\; 0\,,
\end{equation}
where $\tau>0$ and
\begin{equation}
H(\tau) \;=\; \Big\langle \bigl( e^\LLL+e^{-\LLL} \bigr)_{ab}\Big\rangle_{a \prec b \,\wedge\, \tau_{ab}=\tau}
\;=\; 2  e^{-\gamma_E} G_0^R(\tau) + \tfrac12 e^{\gamma_E} D(\tau)\,.
\end{equation}
Verifying the CST-analog of the equation of motion~(\ref{eom}) numerically, we demonstrate in Fig.\ref{fig:diffeq} that the two terms perfectly cancel each other over the entire range of $\tau\in(0,1)$, thus supporting the conjecture that $\DDD$ approximates $\quabla$.

The present construction is manifestly compatible with the Lorentz-invariant spirit of causal set theory, since it is defined entirely in terms of the intrinsic order-theoretic data of a Poisson sprinkling and does not depend on any preferred frame \cite{surya2019causal}. Moreover, although the operator is not local in the strict sense---because all powers of the link matrix contribute---its effective behavior is nevertheless short-ranged: the contributions from higher-order terms are strongly suppressed, and the corresponding averaged kernel rapidly decays at large proper time. In this sense, the construction is best viewed as a Lorentz-invariant, retarded, and effectively short-range candidate inverse kernel whose precise relation to existing causal-set d'Alembertians remains an important open problem.

Therefore, it is not yet clear whether the inverse kernel proposed here can be regarded as a complete  causal-set d'Alembertian. Neither is it established that the operator in Eq. \eqref{DDef} provides, in full generality, the same level of description as the nonlocal retarded Benincasa--Dowker-type d'Alembertians and their higher-dimensional and generalized variants \cite{PhysRevLett.104.181301,dowker2013causal,aslanbeigi2014generalized}, nor as the more recent local constructions that aim to recover the continuum wave operator by combining intrinsically defined timelike and spacelike structures on the causal set \cite{np89-qzjp}. 

At present, our results should therefore be understood more modestly: they show that the inverse kernel naturally associated with the link-based propagator is short-ranged and reproduces the expected behavior on the class of test functions studied here, but further numerical work and, especially, a substantially deeper analytic understanding will be required before any stronger claim can be justified.

\section{Conclusions}

\noindent
In this work, we investigate whether retarded scalar propagators on causal sets can be constructed directly from the link matrix $\LLL$, rather than from the causal matrix $\CCC$. This question is motivated by the view that links encode the local adjacency structure of a causal set and therefore provide a more intrinsic starting point for dynamical constructions. In the standard $1+1$-dimensional causal-set prescription, the massless retarded propagator is usually taken to be proportional to $\CCC$, since this reproduces the correct continuum expectation value after averaging. By contrast, our aim here was to explore whether the same continuum behavior can be recovered from a genuinely link-based construction.

For Poisson sprinklings into $1+1$-dimensional Minkowski spacetime, our results indicate that this is indeed possible. More precisely, within the asymptotic framework developed in Section~III, the requirement that the averaged kernel approaches a constant at large proper time selects the exponential series in the link matrix. This leads us to the proposal that
\begin{equation}
\GGG_0^R = \frac{1}{2} e^{\gamma_E} \exp(\LLL)
\notag
\end{equation}
is a possible candidate for a causal set propagator in the sense that its expectation value tends asymptotically to the correct continuum value. In this sense, the massless retarded propagator can be intrinsically represented in terms of the link structure alone. The numerical results presented here support this conclusion and confirm that its averaged matrix elements rapidly approach the predicted constant.

The same construction extends naturally to the massive case. Replacing continuum convolutions by sums over sprinkled elements yields the expected discrete mass-scattering expansion, and inserting the link-based massless kernel leads to a massive causal-set propagator in good agreement with the continuum retarded propagator after averaging. This confirms that the exponential link construction is compatible with the established path-sum framework and is not limited to the massless case.

Our results also suggest a natural candidate for a corresponding inverse kernel, namely
\begin{equation}
D = 2 e^{-\gamma_E} \exp(-\LLL).
\notag
\end{equation}
The numerical evidence indicates that this operator is short-ranged and reproduces the expected action of the continuum wave operator $\quabla$ on a class of $\tau$-dependent test functions. However, this interpretation should still be considered as preliminary. In particular, we restricted the analysis to test functions that depend exclusively on $\tau$. Moreover, the precise analytic relation between this operator and other existing causal-set d'Alembertians in the literature~\cite{sorkin2011scalar,dowker2013causal,aslanbeigi2014generalized,belenchia2016continuum,np89-qzjp} remains to be clarified.

More broadly, the present work points toward the possibility that scalar propagators on causal sets may admit a unified link-based description, with different dimensions selecting different combinations of powers of the link matrix. Establishing such a framework more systematically and deriving it analytically rather than numerically remains an important problem for future work.

A natural next step will be to investigate whether the propagator construction developed here can be extended beyond flat spacetime to $1+1$-dimensional spacetimes of constant curvature, in particular de Sitter and anti-de Sitter spaces. Since propagators on causal sets in such backgrounds exhibit characteristic curvature-dependent behavior, this provides a particularly instructive setting in which to test the robustness of the present proposal. It will be especially interesting to determine whether the exponential link construction survives unchanged, deforms into a curvature-dependent function of the link matrix, or has to be replaced by a more general link-based kernel.

\appendix
\section{Estimation of the integral expression}
\label{AppendixCalculation}

\begin{figure}[t!]
\includegraphics[width=80mm]{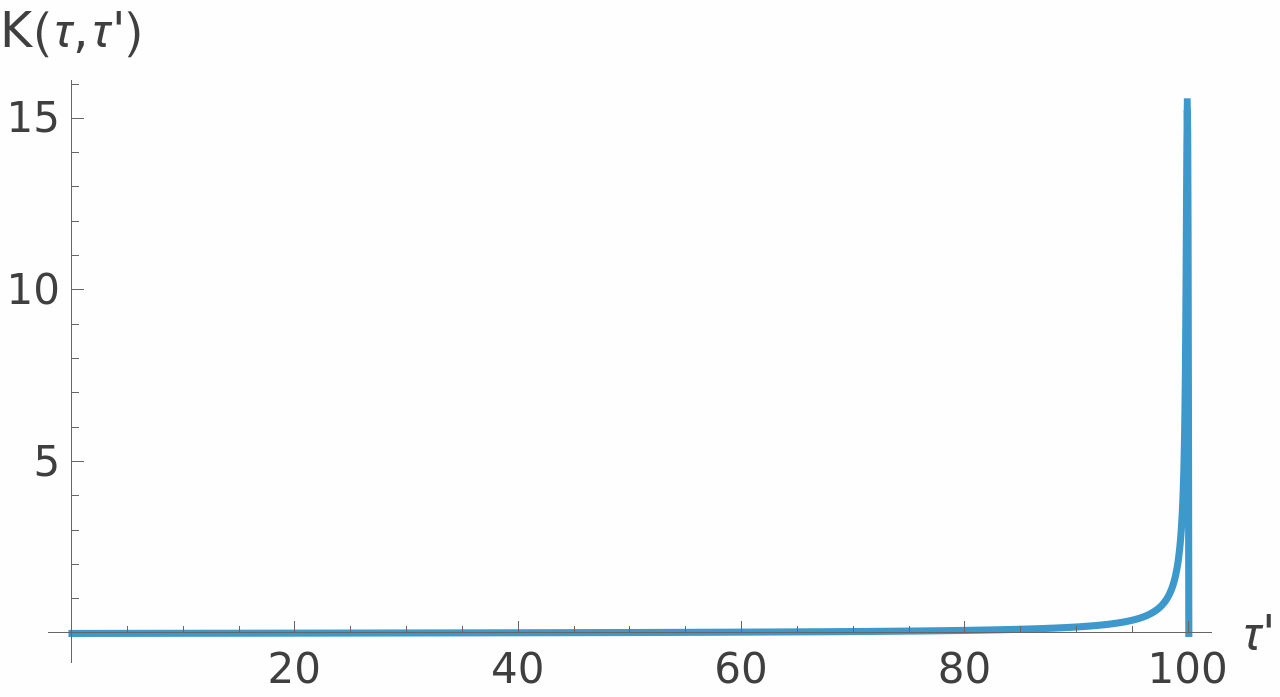}
\caption{Reduced kernel $\tilde K(\tau,\tau')$ for $\tau=100$ and $\rho=100$ as a function of $0<\tau<\tau'$.
\label{fig:peak}
}
\end{figure}

\noindent
In this Appendix, we outline how Eqs.~(\ref{claim1})-(\ref{claim2}) can be proven. The first step is to rewrite the l.h.s. of Eq.~(\ref{claim1}) as
\begin{equation}
\label{HGF}
H(\tau)\;:=\;
\bigl(G \circledast F_1\bigr)(\tau)
\;=\; \int_0^\tau \d\tau'\, \tilde K(\tau,\tau') \, G(\tau')\,,
\end{equation}
where
\begin{equation}
\label{ReducedKernel}
\tilde K(\tau,\tau')\;=\; \int_0^{\tau-\tau'}\d \tau''
\,\,\,K(\tau,\tau',\tau'')\,\,\,
e^{-\frac12 \rho(\tau'')^2}
\end{equation}
is a reduced kernel. Unfortunately, this reduced kernel cannot be computed in closed form. In order to still arrive at a result for large $\tau$, it is helpful to note that the reduced kernel $\tilde K(\tau,\tau')$ exhibits a sharp peak for $\tau' \to \tau$ (see Fig.~\ref{fig:peak}). This suggests splitting the integral in Eq.~(\ref{HGF}) into a \textit{singular} part close to the peak and a \textit{regular} part in the bulk:
\begin{equation}
H(\tau)\;=\;H_{\text{s}}(\tau)+H_{\text{r}}(\tau)\,.
\end{equation}
In both parts, the kernel can be approximated in a different way, allowing the integration to be carried out to the leading order in $\tau$.

\paragraph{Estimation of the singular part\\}
%
In the vicinity of the peak, both $s:=\tau-\tau_1$ and $\tau''\in[0,s]$ are  small. In this situation, the full kernel can be approximated as
\begin{equation}
K(\tau,\tau_1,\tau_2) \approx \frac{2\,\rho\,\tau_2}{\sqrt{s^2-(\tau')^2}}\,.
\end{equation}
Using this approximation one obtains the reduced kernel
\begin{equation}
\label{ksapprox}
\tilde K(\tau,\tau_1)\;\approx\;2 \sqrt{2\rho}\, \,\mathcal F\Bigl( s\sqrt{\tfrac{\rho}{2}} \Bigr)\,
\end{equation}
valid in the vicinity of the peak,
where $\mathcal F(z) \approx e^{-z^2}\int_0^z e^{t^2}\d t$ denotes the Dawson integral. Inserting (\ref{ksapprox}) into (\ref{HGF}) allows us to express the singular contribution in terms of a hypergeometric function as
\begin{equation}
H_\text{s}(\tau)
\;\approx\; G(\tau)\,\int_0^\tau \d\tau_1\, \tilde K(\tau,\tau_1)\;=\; G(\tau)\,
\rho \tau^2 \,
{}_2F_2\Bigl( 1,1;\,\,\tfrac32,2,\,\,-\tfrac{\rho\tau^2}{2} \Bigr)\,.
\end{equation}
In the limit of large sprinkling densities $\rho\to\infty$ the singular contribution can thus be approximated by
\begin{equation}
\label{singularContribution}
H_\text{s}(\tau) \;\approx\;  G(\tau)\,\Bigl(2\ln\tau +\ln\rho + \ln 2 + \gamma_E\Bigr)\,.
\end{equation}
\paragraph{Estimation of the regular part\\}
%
In the remaining integration range away from the peak, the function $F_1(\tau'')=\exp(-\tfrac12 \rho^2(\tau''))$ still forces $\tau''$ to be small, but now the difference $s=\tau-\tau'$ is much larger than $\tau''$. This requires us to use a different type of approximation for the kernel, namely
\begin{equation}
K(\tau,\tau_1,\tau_2) \;\approx\; \frac{4\rho\tau_1\tau_2}{\tau^2-\tau_1^2}\,,
\end{equation}
leading also to a different approximation of the reduced kernel
\begin{equation}
\label{approxReduced}
\tilde K(\tau,\tau_1)\;\approx\;
\frac{4\tau_1}{\tau^2-\tau_1^2}\,.
\end{equation}
This approximation is still singular in the limit $\tau_1\to\tau$. However, above we have already taken the singular contributions into account. Therefore, when integrating over the whole range of $\tau'$ from 0 to $\tau$, we have to carefully eliminate this singularity. This can be achieved by applying a partial fraction decomposition, splitting the integral into two parts
\begin{equation}
\label{fractionDecomposition}
H_{\text{r}}(\tau)\;\approx\;
\frac{2}{\tau}\int_0^\tau \frac{\tau'}{\tau+\tau'}
G(\tau')\d\tau'
\;+\;
\frac{2}{\tau}\int_0^\tau \frac{\tau'}{\tau-\tau'}
G(\tau')\d\tau'\,.
\end{equation}
The first integral in non-divergent. Inserting the assumed power law $G(\tau) \simeq \alpha \, \tau^{-\beta}$ given in~ (\ref{powerlawansatz}), we obtain
\begin{equation}
\frac{2}{\tau}\int_0^\tau \frac{\tau'}{\tau+\tau'}
G(\tau')\d\tau'
 \;=\; G(\tau)
\Bigl( \psi\bigl(\tfrac{3-\beta}{2}\bigr)-\psi\bigl(\tfrac{2-\beta}{2}\bigr) \Bigr)\,,
\end{equation}
where $\psi(z)=\Gamma'(z)/\Gamma(z)$ denotes the digamma function.

The second integral in Eq.~(\ref{fractionDecomposition}) is logarithmically divergent in the limit $\tau_1 \to \tau$. However, since we have already taken the singular contribution at the peak into account, we again have to remove this divergence. This can be done by a suitable regularization in the numerator, i.e.,
\begin{equation}
\frac{2}{\tau}\int_0^\tau \frac{\tau'\,G(\tau')}{\tau-\tau'}
\d\tau'
\;\longrightarrow\;
\frac{2}{\tau}\int_0^\tau \, \d\tau'\,  \frac{\tau' G(\tau')-\tau G(\tau)}{\tau-\tau'}\,.
\end{equation}
This regularizes the integral and allows it to be evaluated:
\begin{equation}
\frac{2}{\tau}\int_0^\tau \, \d\tau'\,  \frac{\tau' G(\tau')-\tau G(\tau)}{\tau-\tau'}
\;=\; G(\tau)\Bigl( -2\gamma_E -2\psi\bigl(2-\beta\bigr)\Bigr)
\end{equation}
Combining the expression for the two integrals in (\ref{fractionDecomposition}), the regular contribution therefore reads
:%
\begin{equation}
\label{regularContribution}
H_\text{r}(\tau)
\;=\; G(\tau)\Bigl(
-2\gamma_E
+ \psi\bigl(\frac{3-\beta}{2}\bigr)
- \psi\bigl(\frac{2-\beta}{2}\bigr)
- 2\psi\bigl(2-\beta\bigr)
\Bigr)
\end{equation}
\paragraph{Combination of the singular and the regular part\\}
%
Combining (\ref{singularContribution}) and (\ref{regularContribution}) and taking into account that $\beta$ actually depends on $\lambda$, we obtain the final result
\begin{equation}
H(\tau) \;=\; 
H_s(\tau) + H_r(\tau) \;=\; 
G_\lambda(\tau) \Bigl( 2 \ln \tau + C_\lambda  \Bigr)
\end{equation}
where
\begin{equation}
C_\lambda \;=\; \ln \rho+\ln 2 - \gamma_E
+ \psi\Bigl(\tfrac{3-\beta(\lambda)}{2}\Bigr)
- \psi\Bigl(\tfrac{2-\beta(\lambda)}{2}\Bigr)
- 2\psi\Bigl(2-\beta(\lambda)\Bigr)
\end{equation}
is a $\tau$-independent constant and where $\psi(z)=\Gamma'(z)/\Gamma(z)$ denotes the digamma function.



\pagebreak
\bibliography{references}

\end{document}